\documentclass[preprint,prl,a4paper,superscriptaddress,showpacs,preprintnumbers]{revtex4}
\usepackage{graphicx}
\newcommand{\GeV}{\,\mbox{GeV}}
\newcommand{\MeV}{\,\mbox{MeV}}

\begin{document}
 
\vspace*{-3\baselineskip}
\resizebox{!}{2.5cm}{\includegraphics{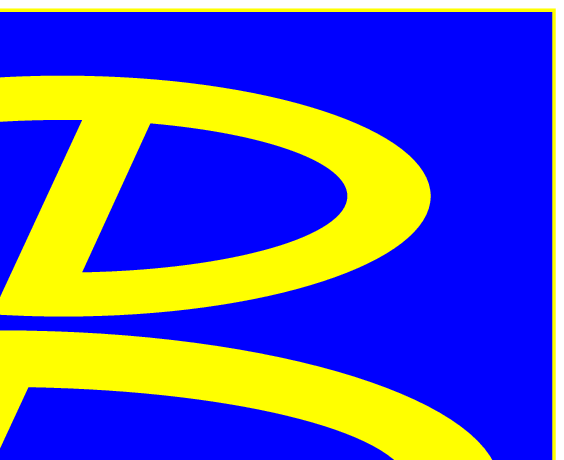}}
\preprint{\vbox{
                \hbox{Belle preprint 2002-39 } 
                \hbox{KEK preprint   2002-127 }
                \hbox{hep-ex/0212052} 
}}

  \title{ \boldmath Observation of
    ${\bar{B^0}\to\Lambda_c^+\bar{p}}$ decay}

\affiliation{Budker Institute of Nuclear Physics, Novosibirsk}
\affiliation{Chiba University, Chiba}
\affiliation{Chuo University, Tokyo}
\affiliation{University of Cincinnati, Cincinnati, Ohio 45221}
\affiliation{University of Frankfurt, Frankfurt}
\affiliation{Gyeongsang National University, Chinju}
\affiliation{University of Hawaii, Honolulu, Hawaii 96822}
\affiliation{High Energy Accelerator Research Organization (KEK), Tsukuba}
\affiliation{Hiroshima Institute of Technology, Hiroshima}
\affiliation{Institute of High Energy Physics, Chinese Academy of Sciences, Beijing}
\affiliation{Institute of High Energy Physics, Vienna}
\affiliation{Institute for Theoretical and Experimental Physics, Moscow}
\affiliation{J. Stefan Institute, Ljubljana}
\affiliation{Kanagawa University, Yokohama}
\affiliation{Korea University, Seoul}
\affiliation{Kyoto University, Kyoto}
\affiliation{Kyungpook National University, Taegu}
\affiliation{Institut de Physique des Hautes \'Energies, Universit\'e de Lausanne, Lausanne}
\affiliation{University of Ljubljana, Ljubljana}
\affiliation{University of Maribor, Maribor}
\affiliation{University of Melbourne, Victoria}
\affiliation{Nagoya University, Nagoya}
\affiliation{Nara Women's University, Nara}
\affiliation{National Lien-Ho Institute of Technology, Miao Li}
\affiliation{National Taiwan University, Taipei}
\affiliation{H. Niewodniczanski Institute of Nuclear Physics, Krakow}
\affiliation{Nihon Dental College, Niigata}
\affiliation{Niigata University, Niigata}
\affiliation{Osaka City University, Osaka}
\affiliation{Osaka University, Osaka}
\affiliation{Panjab University, Chandigarh}
\affiliation{Peking University, Beijing}
\affiliation{RIKEN BNL Research Center, Upton, New York 11973}
\affiliation{Saga University, Saga}
\affiliation{University of Science and Technology of China, Hefei}
\affiliation{Seoul National University, Seoul}
\affiliation{Sungkyunkwan University, Suwon}
\affiliation{University of Sydney, Sydney NSW}
\affiliation{Tata Institute of Fundamental Research, Bombay}
\affiliation{Toho University, Funabashi}
\affiliation{Tohoku Gakuin University, Tagajo}
\affiliation{Tohoku University, Sendai}
\affiliation{University of Tokyo, Tokyo}
\affiliation{Tokyo Institute of Technology, Tokyo}
\affiliation{Tokyo Metropolitan University, Tokyo}
\affiliation{Tokyo University of Agriculture and Technology, Tokyo}
\affiliation{Toyama National College of Maritime Technology, Toyama}
\affiliation{University of Tsukuba, Tsukuba}
\affiliation{Utkal University, Bhubaneswer}
\affiliation{Virginia Polytechnic Institute and State University, Blacksburg, Virginia 24061}
\affiliation{Yokkaichi University, Yokkaichi}
\affiliation{Yonsei University, Seoul}
  \author{N.~Gabyshev}\affiliation{High Energy Accelerator Research Organization (KEK), Tsukuba} 
  \author{H.~Kichimi}\affiliation{High Energy Accelerator Research Organization (KEK), Tsukuba} 
  \author{K.~Abe}\affiliation{High Energy Accelerator Research Organization (KEK), Tsukuba} 
  \author{K.~Abe}\affiliation{Tohoku Gakuin University, Tagajo} 
  \author{R.~Abe}\affiliation{Niigata University, Niigata} 
  \author{T.~Abe}\affiliation{Tohoku University, Sendai} 
  \author{I.~Adachi}\affiliation{High Energy Accelerator Research Organization (KEK), Tsukuba} 
  \author{H.~Aihara}\affiliation{University of Tokyo, Tokyo} 
  \author{M.~Akatsu}\affiliation{Nagoya University, Nagoya} 
  \author{Y.~Asano}\affiliation{University of Tsukuba, Tsukuba} 
% ...
\collaboration{The Belle Collaboration}

  \begin{abstract}
  We report the measurement of the charmed baryonic decay 
  $\bar{B^0}\to\Lambda_c^+\bar{p}$ with a branching fraction of 
  $(2.19^{+0.56}_{-0.49}\pm0.32\pm0.57)\times10^{-5}$
  and a statistical significance of $5.8\,\sigma$.
  The errors are statistical, systematic,
  and the error of the $\Lambda_c^+\to{p}K^-\pi^+$ decay branching fraction. 
  This is the first observation of a two-body baryonic $B$ decay.
  The analysis is based on 78.2\,fb$^{-1}$ of data accumulated at the
  $\Upsilon(4S)$ resonance with the Belle detector at the KEKB
  asymmetric $e^+ e^-$ collider.   

  \end{abstract}

  \pacs{13.25.Hw, 14.20.Lq}
 
  \maketitle

  Although the four- and three-body baryonic $B$ decays
  $\bar{B^0}\to\Lambda_c^+\bar{p}\pi^+\pi^-$ and
  ${B^-}\to\Lambda_c^+\bar{p}\pi^-$ 
  are experimentally well established\,\mbox{\cite{cleo-lamc,blamc}},
  there has been, until now, no reported observation
  of any two-body mode, such as
  $\bar{B^0}\to\Lambda_c^+\bar{p}$.
  In a previous Belle analysis, based on a 29.1\,fb$^{-1}$ data
  sample\,\cite{blamc}, we obtained the following branching fractions 
  for \mbox{four-,} three- and two-body decays:
\begin{center}
${\cal{B}}(\bar{B^0}\to\Lambda_c^+\bar{p}\pi^+\pi^-) = 
  (11.04^{+1.22}_{-1.17}\pm1.98\pm2.87)\times10^{-4}$, \\
${\cal{B}}({B^-}\to\Lambda_c^+\bar{p}\pi^-)          = 
  (1.87^{+0.43}_{-0.40}\pm0.28\pm0.49)\times10^{-4}$, \\
${\cal{B}}(\bar{B^0}\to\Lambda_c^+\bar{p}) < 
{0.31\times10^{-4}}\ \mbox{(90\% confidence level)}$. \\
\end{center}
  The measured branching fractions decrease rapidly
  with decreasing decay multiplicity.
  This suppression of lower multiplicity decays is a key issue in
  the understanding of the mechanism behind charmed baryonic $B$ decays.

  There are several different theoretical calculations for the 
  $\bar{B^0}\to\Lambda_c^+\bar{p}$  branching fraction,
  based on a diquark model\,\cite{duquark}, a QCD sum rule
  model\,\cite{qcd_sum_rule} and pole models\,\cite{pole,bag}.
  They differ by an order of magnitude. 
  Thus, a  measurement of the two-body decay 
  $\bar{B^0}\to\Lambda_c^+\bar{p}$ branching fraction 
  would distinguish between these different theoretical approaches
  and provide important insight into the underlying physics.

  In this paper we report the first observation of the 
  two-body decay $\bar{B^0}\to\Lambda_c^+\bar{p}$.
  The analysis is based on a data sample of 78.2\,fb$^{-1}$ accumulated
  at the $\Upsilon(4S)$ resonance with the Belle detector at the KEKB
  8\GeV\,$e^-$ on 3.5\GeV\,$e^+$ asymmetric collider.

  The Belle detector is a large-solid-angle magnetic
  spectrometer that consists of a three-layer silicon vertex
  detector (SVD), a 50-layer cylindrical drift chamber (CDC),
  a mosaic of aerogel threshold \v{C}erenkov counters (ACC),
  a barrel-like array of time-of-flight scintillation counters (TOF),
  and an array of  CsI(Tl) crystals (ECL) located inside a
  superconducting solenoidal coil that provides a 1.5\,T magnetic field.
  An iron flux return located outside the coil
  is instrumented to detect muons and $K_L$ mesons (KLM).
  The detector is described in detail elsewhere\ \cite{belle}.
  We use a GEANT based Monte Carlo (MC) simulation to model the
  response of the detector and determine its acceptance\ \cite{sim}.

  We detect the $\Lambda_c^+$ via the $\Lambda_c^+\to{pK^-\pi^+}$ 
  decay channel.
  Inclusion of charge conjugate states is implicit unless otherwise
  stated. 
  The particle identification (PID) information from the CDC, ACC and
  TOF is used to construct likelihood functions $L_p$, $L_K$ and
  $L_{\pi}$ for the proton, kaon and pion assignment for all charged
  tracks, respectively.  
  Likelihood ratios $LR(A/B)=L_A/(L_A+L_B)$ are required to be
  greater than 0.6 to identify a particle from ${\Lambda_c^+}$ decay
  as type $A$, where $B$ denotes the other two possible assignments among 
  kaon, pion or proton.
  In order to maintain high efficiency for the high momentum prompt
  antiproton that comes directly from the primary $\bar{B^0}$ meson
  decay, we rely more heavily on the  kinematic reconstruction and
  loosen the PID requirement to $LR(A/B)>0.2$, which
  improves the efficiency by a factor of about 20\%.
  Electron and muon candidates are removed if their combined 
  likelihood ratios from the ECL, CDC and KLM information are greater
  than 0.95. 
  A ${\Lambda_c^+}$ candidate is selected if the invariant mass
  $M(pK^-\pi^+)$ is within $0.010\GeV/c^2$ ($2.5\,\sigma$) 
  of the $2.285\GeV/c^2$ ${\Lambda_c^+}$ mass.
  A ${\Lambda_c^+}$ mass constrained fit is carried out at the
  reconstructed ${\Lambda_c^+}$ decay vertex to remove background 
  including secondary particles from $\Lambda$ or $K_S$ decay.

  The $\bar{B^0}\to\Lambda_c^+\bar{p}$ events are identified 
  by their energy difference $\Delta{E}=(\sum E_i)-E_{\rm{beam}}$, and 
  the beam-energy constrained mass 
  $M_{\rm{bc}}=\sqrt{E^2_{\rm{beam}}-(\sum\vec{p}_i)^2}$, where  
  $E_{\rm{beam}}$ is the beam energy, and $\vec{p}_i$ and $E_i$ are the
  three-momenta and energies of the $B$ meson decay products,
  all defined in the center-of-mass system of the
  $e^+e^-$ collision.
  We select events with $M_{\rm{bc}}>5.20$\GeV/$c^2$ and
  $|\Delta{E}|<0.20$\GeV. 
  The prompt antiproton track and the virtual ${\Lambda_c^+}$ track 
  are required to form a common $B$ mass constrained decay vertex.
  To suppress continuum background, we impose requirements on
  event-shape variables.
  We require $|\cos\theta_{\rm{thr}}|<0.80$, where $\theta_{\rm{thr}}$
  is the angle between the thrust axis of the $B$ candidate tracks and
  that of the other tracks.
  This requirement eliminates 80\% of the continuum background and
  retains 80\% of the signal events. 
  We also require $R_2<0.35$, where $R_2$ is the ratio of the second
  to the zeroth Fox-Wolfram moments\,\cite{fox_wolfram}.
  This requirement rejects 50\% of the remaining continuum background
  and retains 90\% of the signal.
%
%
  If there are multiple candidates in an event, the candidate with
  the best $\chi^2_{B}$ for the $B$ vertex fit is selected.
  
  \begin{figure*}[htb]
    \begin{tabular}{cc}
      \begin{minipage}{0.5\textwidth}
	\includegraphics[width=0.9\textwidth]{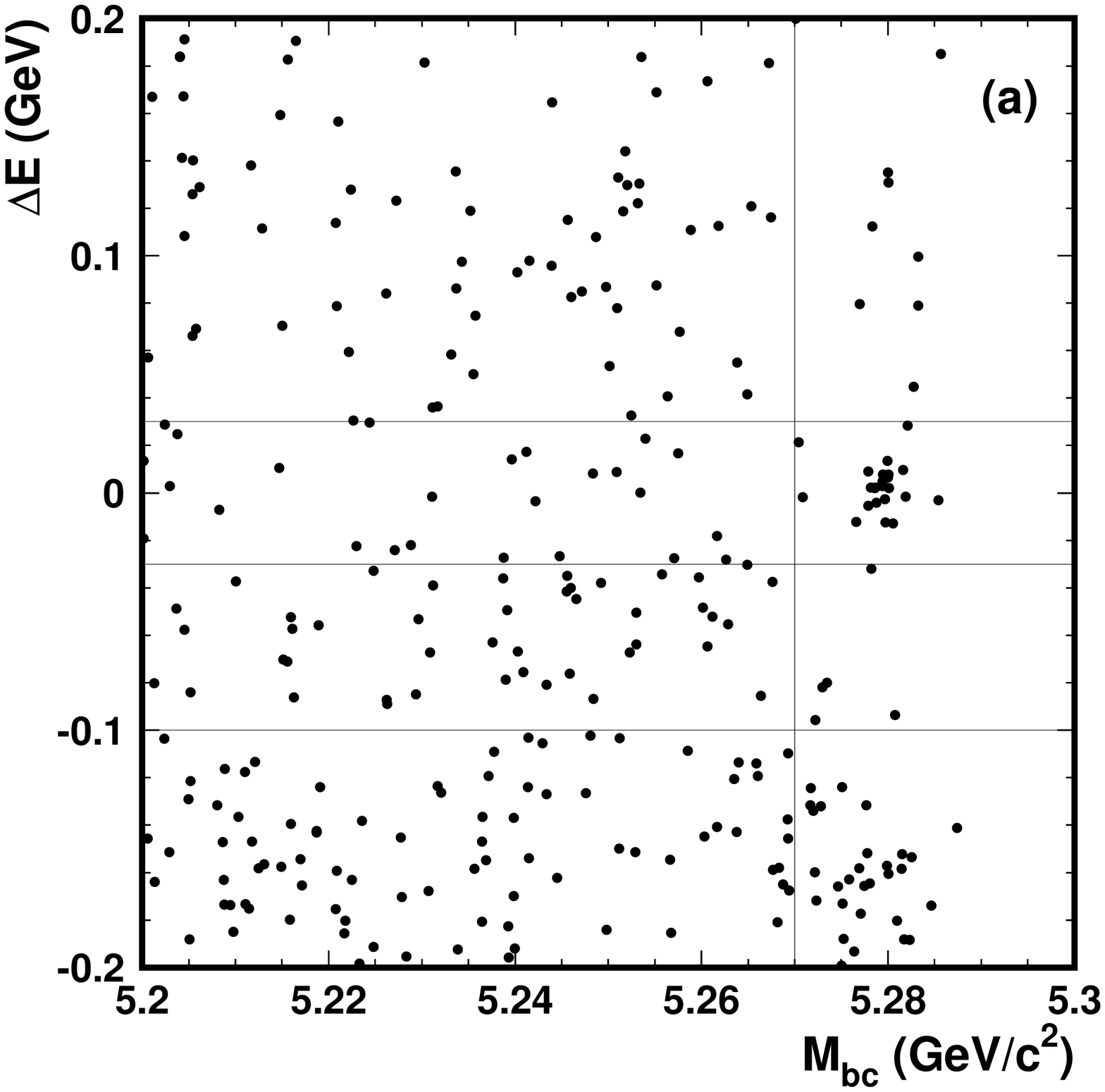} 
      \end{minipage}
      &
      \begin{minipage}{0.5\textwidth}
	\includegraphics[width=0.9\textwidth]{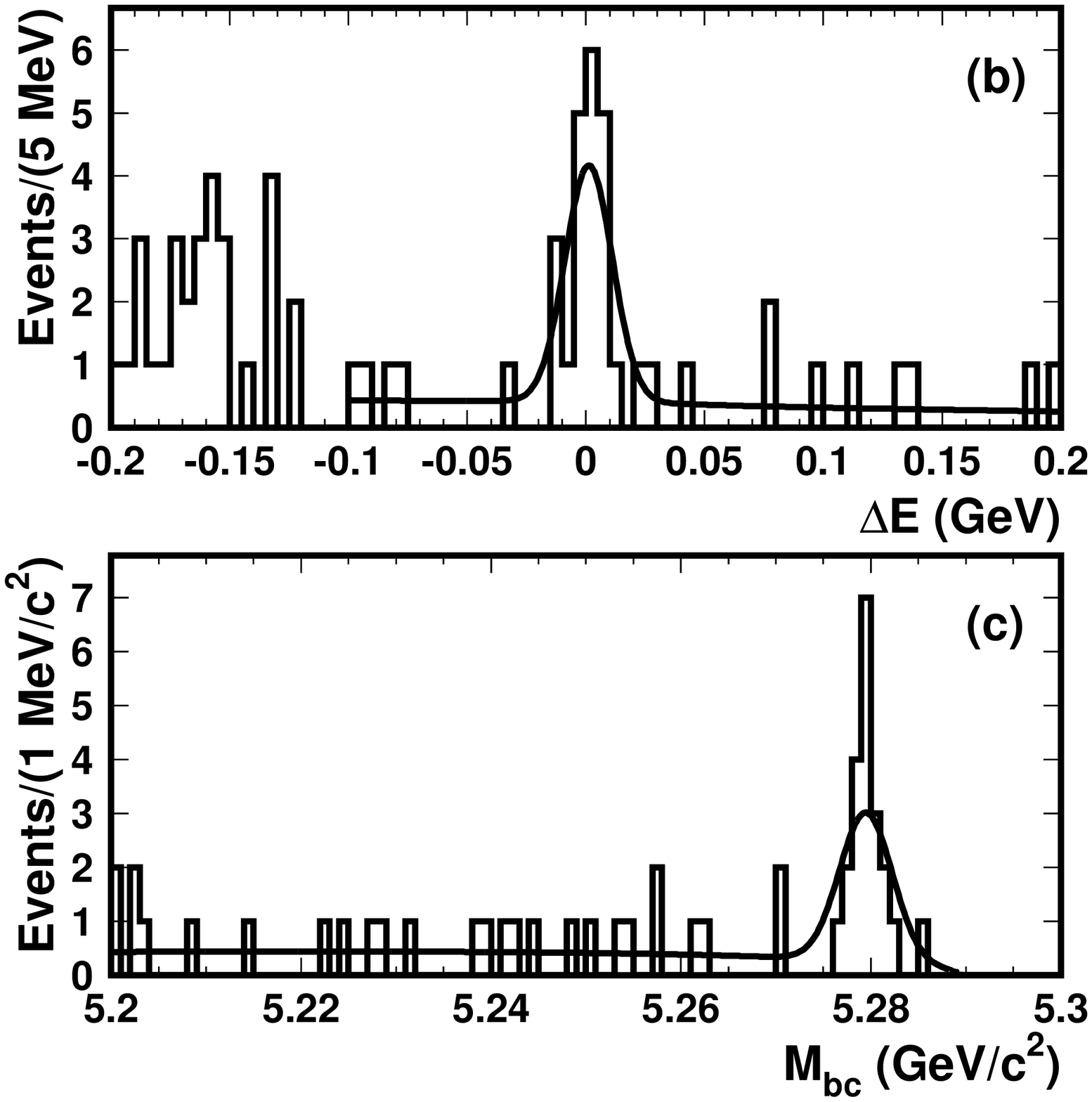} 
      \end{minipage}
      \\
    \end{tabular}
    \caption{ Candidate $\bar{B^0}\to\Lambda_c^+\bar{p}$ events:
      (a)\,scatter plot of $\Delta{E}$ versus $M_{\rm{bc}}$,
      (b)\,$\Delta{E}$ distribution for $M_{\rm{bc}}>5.270\GeV/c^2$ and
      (c)\,$M_{\rm{bc}}$ distribution for $|\Delta{E}|<0.030\GeV$.
      The curves indicate the result of a two-dimensional fit.
    }
    \label{fig:explamc1}
  \end{figure*}
  Figure~\ref{fig:explamc1} shows a scatter plot of $\Delta{E}$ 
  versus $M_{\rm{bc}}$ and their projections for selected events.  
  The $\Delta{E}$ projection is shown for $M_{\rm{bc}}>5.270\GeV/c^2$ and 
  the $M_{\rm{bc}}$ projection is shown for $|\Delta{E}|<0.030\GeV$. 
    The widths determined from single Gaussian fits to signal MC
    events are $2.7\,\MeV/c^2$ and $10.3\,\MeV$ for $M_{\rm{bc}}$
    and $\Delta E$, respectively.
  A two-dimensional binned maximum likelihood fit is performed to
  determine the signal yield. 
  For this fit, the 
  $\Delta{E}$ distribution is represented by 
  a double Gaussian for the signal plus a first order polynomial
  for the background, and the $M_{\rm{bc}}$ distribution is represented 
  by a single Gaussian for the signal plus the ARGUS
  function\,\cite{argus_function} for the background. 
  The region $\Delta{E}<-0.1\GeV$ is excluded 
  from the fit to avoid feed down from modes including extra pions,
  which produces the bump structure observed in the region
  $\Delta{E}\leq-0.15\GeV$.  
  In the fit, the signal shape parameters are fixed to the values
  fitted to the signal MC, 
  and the signal yield and the background parameters are allowed
  to float.
  The curves in Figure~\ref{fig:explamc1}(b) and (c) 
  indicate the results of this two-dimensional fit. 

  The signal peak positions determined from 
  fits to the data, 
  $(5279.5\pm0.3)\MeV/c^2$ for $M_{\rm{bc}}$  
  and $(0.9\pm1.8)\MeV$ for $\Delta{E}$,
  are consistent with the world average $B^0$ mass\,\cite{pdg2002}
  and zero, respectively.
  When we use single Gaussians
  for $M_{\rm bc}$ and $\Delta E$ signal functions
  and fit with the widths as free
  parameters, the fitted values in the data are found to be   
  $(1.3\pm0.3)\MeV/c^2$ 
  and $(6.9\pm1.5)\MeV$, respectively, which are
  narrower than those determined with the signal
  MC. 
  The probability of obtaining such narrow widths is ${\cal{O}}(1\%)$
  and is attributed to a statistical fluctuation.
  We also investigate the decays
  $\bar{B^0}\to\Lambda_c^+ \bar{p} \pi^+ \pi^-$,
  $B^-\to\Lambda_c^+ \bar{p} \pi^-$ and
  $\bar{B^0}{\to}J/\psi\bar{K}^{*}(892)^0$, $J/\psi\to{p\bar{p}}$ as
  control samples, 
  and find that for these modes $M_{\rm{bc}}$ and $\Delta{E}$ widths
  are consistent between the data and signal MC. 
  The effect of the narrow widths to the signal yield is investigated
  by applying fits where single Gaussians with widths allowed to float
  are used for the signal shapes.
  The difference in the fitted yields is taken into account in a
  systematic error as discussed below.
 
  From the fit we obtain $19.6^{+5.0}_{-4.4}$ signal events.
  From separate fits to the charge conjugate modes
  we obtain signal yields of $6.4^{+3.0}_{-2.4}$
  and $13.3^{+4.2}_{-3.5}$ events 
  for $\bar{B^0}\to\Lambda_c^+\bar{p}$ and
  $B^0\to\bar{\Lambda}_c^-{p}$, respectively. 
  These are consistent within statistical errors.

  The branching fraction is calculated as $N_{\rm S}/(\varepsilon
  \times {N_{B\bar{B}}} \times {\cal{B}}(\Lambda_c^+\to{pK^-\pi^+}))$,
  using the measured signal yield $N_{\rm S}$ and
  the decay branching fraction
  ${\cal{B}}(\Lambda_c^+\to{pK^-\pi^+})=(5.0\pm1.3)\%$\,\cite{pdg2002}.
  The detection efficiency $\varepsilon$ is evaluated to be
  21.1\% from the signal MC.
  The number of $B\bar{B}$ pairs $N_{B\bar{B}}$ is
  $(85.0\pm0.5)\times{10^6}$.
  The fractions of charged and neutral $B$ mesons are assumed to be the
  same.

  We obtain a branching fraction of
  $$
  {\cal{B}}(\bar{B^0}\to\Lambda_c^+\bar{p})=
  (2.19^{+0.56}_{-0.49}\pm0.32\pm0.57)\times10^{-5}, 
  $$
  where the first and the second errors are statistical and systematic,
  respectively.
  The last error of 26\% is due to 
  uncertainty in the branching fraction
  ${\cal{B}}(\Lambda_c^+\to{p}K^-\pi^+)$. 

  The total systematic error of 14.8\% is determined as follows. 
  The tracking systematic error is estimated to be 8\% in total, 
  assuming a correlated systematic error of 2\% per charged track,
  based on tracking efficiency studies with $\eta \to
  \gamma\gamma$ and $\eta \to \pi^+\pi^-\pi^0$ samples. 
  The PID systematic error is 10\% in total,
  assuming a correlated systematic error of 3\% per proton and 2\% per
  pion or kaon, based on studies with a $\Lambda \to p \pi^-$ sample
  for protons; and with a $D^{*+} \to D^0\pi^+$, $D^0 \to K^-\pi^+$
  sample for kaons and pions. 
  The systematic error in the fitting procedure and signal shape is
  estimated to be 7.3\%,  
  which is half of the maximum
  deviation in the branching fractions
  obtained with various modifications to the fitting functions: 
  with a single or a double Gaussian for the $\Delta{E}$ signal, with the
  widths and means for both $M_{\rm{bc}}$ and $\Delta{E}$ signals
  fixed to MC determined values or fitted in the data.
  Finally, the systematic error in the detection efficiency due to 
  MC statistics is 1.3\%.

  \begin{figure}[htb]
    \begin{minipage}{0.5\textwidth}
      \includegraphics[width=0.9\textwidth,bb=0 0 567
	280]{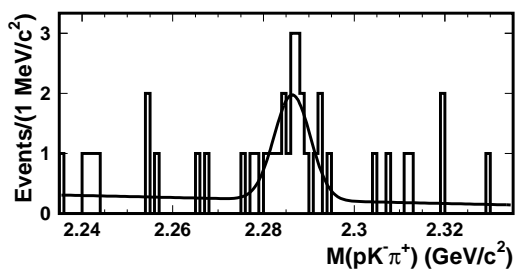}
    \end{minipage}
    \caption{
      Invariant mass $M(pK^-\pi^+)$ distribution for 
      $\bar{B^0}\to\Lambda_c^+\bar{p}$ candidates in the $B$ signal
      region. 
    }
    \label{fig:lcmass}
  \end{figure}
  
  Figure~\ref{fig:lcmass} shows the invariant mass distribution
  $M(pK^-\pi^+)$ for $B$ candidates in the signal region
  $|\Delta{E}|<0.030\GeV$ and $M_{\rm{bc}}>5.27$\GeV/$c^2$. 
  The curve indicates a fit result with a Gaussian over linear background.
  The fitted width of $(4.0\pm1.0)\MeV/c^2$ and
  the fitted mean of $(2286.4\pm1.3)\MeV/c^2$ 
  are consistent with values obtained from fits to signal MC events, 
  which are generated assuming the world average 
  $\Lambda_c^+$ mass\,\cite{pdg2002}.  
  We obtain a ${\Lambda_c^+}$ yield of $17.5^{+5.2}_{-4.6}$ events,
  consistent with the $B$ signal yield mentioned above.

  We consider a contribution in the $\bar{B^0}\to\Lambda_c^+\bar{p}$
  signal yield from other $B$ decays, which gives an 
  uniform distribution in the $\Lambda_c^+$ invariant mass. 
  We analyze the $\Lambda_c^+$ sideband 
  $0.015<|M(pK^-\pi^+)-M_{\Lambda_c^+}|<0.050\GeV/c^2$,
  and obtain a $B$ signal yield of $1.2^{+3.2}_{-2.4}$ events,
  which is consistent with expectation of $1.4\pm0.4$ events 
  from the $\bar{B^0}\to\Lambda_c^+\bar{p}$ decay
  MC, assuming our observed branching fraction.
  From this we estimate the other $B$ decay contribution of
  $(-0.1^{+0.9}_{-0.7})$ events in the $\Lambda_c^+$ signal region, 
  which is negligibly small. 
  From a simultaneous fit of the $\bar{B^0}\to\Lambda_c^+\bar{p}$ signal
  yield and the other $B$ decay contribution in the $\Lambda_c^+$ signal 
  and sideband regions, we obtain a statistical significance of $5.8\,\sigma$.
  The significance is calculated as
  $\sqrt{-2\ln({\cal{L}}_0/{\cal{L}}_{max})}$, where ${\cal{L}}_{max}$
  and ${\cal{L}}_0$ denote the maximum likelihoods with the fitted signal
  yield and with the yield fixed at zero, respectively.

  We investigate the $\bar{B^0}\to\Lambda_c^+\bar{p}$ decay in MC with all
  known $\Lambda_c^+$ decay modes\,\cite{pdg2002}. 
  The overall detection efficiency for the $\Lambda_c^+{\to}pK^-\pi^+$
  final state, including intermediate resonances, is found to be
  consistent with that calculated 
  for the non-resonant $\Lambda_c^+{\to}pK^-\pi^+$ decay alone.
  The other $\Lambda_c^+$ decays show no peaking
  structure in the $\Delta{E}$ and $M_{\rm{bc}}$
  distributions. 

  Finally, the overall systematics are checked by
  analyzing the decay mode 
  $\bar{B^0}{\to}J/\psi\bar{K}^{*}(892)^0$ observed as
  $J/\psi\to{p\bar{p}}$ and $\bar{K}^{*}(892)^0\to K^-\pi^+$,
  which contains the same final state particles as the mode under study.
  A similar analysis procedure is applied, 
  except for the $\Lambda_c^+$ vertex fit.
  A $J/\psi$ is tagged if 
  the measured invariant mass $M(p\bar{p})$
  is within $0.019\GeV/c^2$ of 
  the $J/\psi$ mass.
  A $\bar{K}^{*}(892)^0$ is tagged if 
  the measured invariant mass $M(K^-\pi^+)$
  is within $0.2\GeV/c^2$ of 
  the ${K}^{*}(892)^0$ mass.
  Vertex fits are also carried out at the $J/\psi$ and $B$ vertices.
  We obtain a signal of $23.7^{+5.6}_{-4.9}$ events and 
  a branching fraction of 
  ${\cal B}(\bar{B^0}{\to}J/\psi\bar{K}^{*}(892)^0)=
    (1.02^{+0.24}_{-0.21}\pm0.14)\times10^{-3}$,
  which is consistent with previous
  measurements\,\cite{pdg2002,jpsikst}.

  In summary, 
  we report the measurement of the charmed baryonic $B$ decay 
  $\bar{B^0}\to\Lambda_c^+\bar{p}$ with 
  a branching fraction of 
  $(2.19^{+0.56}_{-0.49}\pm0.32\pm0.57)\times10^{-5}$
  and a statistical significance of $5.8\,\sigma$.
  This is the first observation of a two-body baryonic $B$ decay.
  The branching fraction is found to be about an 
  order-of-magnitude smaller than
  that of the three-body decay $B^-\to\Lambda_c^+\bar{p}\pi^-$. 
  This suppression is a unique feature of two-body baryonic decays;
  in contrast,  the two- and three-body  mesonic $B$ decays 
  are comparable.
  A pole model\,\cite{bag} predicts a value of the branching fraction 
  of $\leq(1.1-3.1)\times 10^{-5}$ for 
  $\bar{B^0}\to\Lambda_c^+\bar{p}$, which is consistent with
  our measurement, while
  the other models\,\cite{duquark,qcd_sum_rule,pole} give substantially
  larger values.
  Charmless baryonic two-body decays
  are expected to be suppressed by an additional factor of 
  ${\vert}V_{ub}/V_{cb}\vert^2$\,\cite{pdg2002}.
  The result reported here implies that their branching fractions
  should not be much above the $10^{-7}$ level, which are consistent
  with the present upper limits\,\cite{belle_charmless_two_body}.  

  \begin{acknowledgments}
    We wish to thank the KEKB accelerator group for their excellent
    operation of the KEKB accelerator.
    We acknowledge support from the Ministry of Education,
    Culture, Sports, Science, and Technology of Japan
    and the Japan Society for the Promotion of Science;
    the Australian Research Council
    and the Australian Department of Industry, Science and Resources;
    National Science Foundation of China under contract No.\,10175071;
    the Department of Science and Technology of India;
    the BK21 program of the Ministry of Education of Korea
    and the CHEP SRC program of the Korea Science and Engineering Foundation;
    the Polish State Committee for Scientific Research
    under contract No.\,2P03B 17017;
    the Ministry of Science and Technology of Russian Federation;
    the Ministry of Education, Science and Sport of Slovenia;
    the National Science Council and the Ministry of Education of Taiwan
    and the U.S.\ Department of Energy.
  \end{acknowledgments}


\end{document}